\documentclass{epl}

\usepackage{epsfig}

\title{Barrier crossing of  semiflexible polymers} 
\shorttitle{Barrier crossing}

\author{P. Kraikivski, R. Lipowsky and J. Kierfeld}
\institute{Max-Planck-Institut f{\"ur} Kolloid- und
  Grenzfl{\"a}chenforschung, 14424 Potsdam, Germany}


\newcommand{\by}{\bar{y}}

\begin{document}

\maketitle

\begin{abstract}
  We consider the motion of  semiflexible polymers
  in  double-well potentials. We calculate
  shape, energy, and effective diffusion constant of kink excitations,
  and in particular their dependence on  the bending rigidity of the
  semiflexible polymer.
  For symmetric potentials, the
  kink motion is purely diffusive whereas kink motion becomes
  directed in the presence of a driving force on the polymer.  
  We determine the average velocity of  the
  semiflexible polymer based on the kink dynamics. The Kramers escape
  over the potential barriers proceeds by nucleation and diffusive
  motion of kink-antikink pairs, the relaxation to the straight
  configuration by annihilation of kink-antikink pairs.
  Our results apply to the activated motion of
  biopolymers such as DNA and actin filaments or synthetic
  polyelectrolytes on structured substrates.
\end{abstract}

\section{Introduction}
The Kramers problem \cite{Kramers} of thermally activated escape of an
object over a potential barrier is one of the central problems of
stochastic dynamics.  It has been extensively studied not only 
for point particles \cite{Haenggi} but also for 
extended objects such as elastic strings which occur
 in a variety of contexts in
condensed matter physics such as dislocation motion in crystals
\cite{dislocation}, motion of flux lines in type-II
superconductors \cite{vortex}, or charge-density waves \cite{CDW}.
Elastic strings activate over potential barriers by nucleation and
subsequent separation of soliton-antisoliton pairs which are
localized  kink excitations \cite{BL79,CKBT80}.
An analogous problem is the activated motion of a flexible
polymer over a potential barrier \cite{sebastian}.

However, the thermally activated escape of a semiflexible polymer,
which is a filament   governed by its bending energy rather than 
entropic elasticity or tension, remained an open question that we want
to address in this paper.
Semiflexible polymers such as DNA or actin filaments
have a large bending stiffness and, thus, a large persistence length, $L_p$.
On scales exceeding $L_p$,  the orientational order of the
polymer segments decays exponentially, and the polymer effectively
behaves as a flexible chain with a segment size set by $L_p$. 
In contrast, on
length scales which are small compared to $L_p$,   the bending
energy  of the semiflexible polymer 
strongly affects the  behaviour of the polymer.
The persistence lengths of the most 
prominent biopolymers range from $50 {\rm nm}$  for
DNA \cite{taylor90},  to  the
$10\mu{\rm m}$-range for actin \cite{gittes93,kaes94}
 or even up to  the ${\rm mm}$-range for
microtubules \cite{gittes93} and becomes  
comparable to typical  contour lengths of  these polymers.
Whereas the adsorption of such semiflexible polymers onto homogeneous
adhesive surfaces has been studied previously in
\cite{NJ99,GB89,KL03},
much less is known about the behaviour of a semiflexible
polymer adsorbed on a {\em structured} surface.

\begin{figure}[h]
\begin{center}
  \epsfig{file=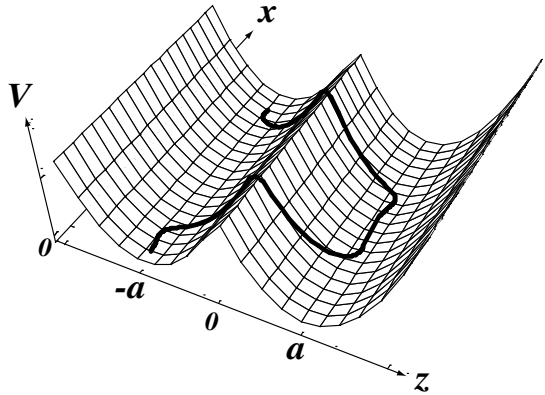,width=0.5\textwidth}
\caption{\label{potential}
Typical conformation of a semiflexible polymer (thick line) with 
a kink-antikink pair in a double-well potential $V$ which 
depends on the coordinate $z$ and is independent of the coordinate $x$. 
}
\end{center}
\end{figure}

In this article we focus on the escape of a  semiflexible polymer over
a translationally invariant potential 
barrier as shown in fig.~\ref{potential}, which can 
be realized on chemically or lithographically structured surfaces.
The behaviour of semiflexible biopolymers on such structured substrates
is of interest, e.g.,  for electrophoresis
applications~\cite{electrophoresis}. 
Another important class of semiflexible polymers are 
synthetic polyelectrolytes, whose  self-assembly and dynamic
behaviour on 
structured substrates has only been studied recently \cite{kurth02}.
In this article we consider homogeneous driving forces 
across the potential barriers as they can be easily realized on  
structured substrates by electric fields for charged polymers as in 
 electrophoresis or  by hydrodynamic flow.
Alternatively, escape over a barrier can be driven by entropic forces
arising from  asymmetric shapes of the potential valleys \cite{CM01}.

Our main results are as follows. 
As for flexible polymers,   the activated dynamics of semiflexible
polymers is governed by the nucleation of localized kink-like
excitations  shown in fig.~\ref{potential}.
We find, however,  that the  activated dynamics 
of  semiflexible polymers is  different from that of flexible
polymers as kink properties are not governed by entropic elasticity of
the polymer chain but rather by the bending energy of the 
semiflexible polymer.
This enables us  to determine the persistence 
length from kink-properties. Furthermore, we  calculate 
time scales for barrier crossing and the  mean
velocity of the semiflexible polymer for all regimes of driving
forces: (i) nucleation and purely diffusive motion of single kinks
(ii) nucleation and driven diffusive motion of single kinks and (iii)
for large driving force dynamic equilibrium between nucleation and
recombination in a kink ensemble.

\section{Model}
We consider the dynamics of a semiflexible polymer in 1+1 dimensions
in a  double-well potential that is translationally invariant in one
direction, say the $x$-axis as in 
 fig.~\ref{potential}.  The semiflexible  polymer 
has a  bending rigidity
$\kappa$ and persistence length $L_p = 2\kappa/T$ where $T$ is the
temperature in energy units. 
We focus on the regime where the  potential is sufficiently strong 
that the  semiflexible
polymer is oriented along the  $x$-axis  and   can be 
parameterized by displacements $z(x)$ perpendicular to the $x$-axis
with $-L/2 < x < L/2$, where $L$ is the projected length of
polymer.  The Hamiltonian of the semiflexible
polymer is given by
\begin{equation}
{\cal H} = \int_{-L/2}^{L/2} dx
  \left[  \frac{\kappa}{2}   
  \left(\partial_x^2z\right)^2
+ V(z)
    \right] 
~,
\label{hamil}
\end{equation}
i.e., the sum  of its bending and potential energy.
We consider a piecewise harmonic double-well potential
$V(z)= \frac{1}{2} V_{0}(|z|-a)^2 -Fz$ that is independent of $x$ and
thus translationally invariant in the $x$-direction, 
where  $V_0$ is the depth of the potential and $F$ an external
driving force density. Below the critical force $F_c \equiv aV_0$ 
the potential has  two minima at $z_{min}^{\pm}=\pm a+F/V_0$.
The Hamiltonian (\ref{hamil}) can be made dimensionless by 
measuring energies in units of a characteristic energy  
$E_{sc}=a^2 \kappa^{1/4} V_0^{3/4}$, the $x$-coordinate
 in units of a
characteristic length  $x_{sc} =(\kappa/V_0)^{1/4}$ and the
$z$-coordinate  in units of $a$.

We consider an overdamped dynamics of  the semiflexible polymer with an
equation of motion 
\begin{equation}
 \gamma \partial_t z
  = -\frac{\delta {\cal H}}{\delta z} + \zeta(x,t)
   = - \kappa  
  \partial_x^{4} z
    -V'(z) +  \zeta(x,t)
\label{EOM}
\end{equation}
 where $\gamma$ is the damping constant and $\zeta$ is a 
Gaussian distributed thermal random force
with $\langle \zeta \rangle=0$ and
$\langle \zeta(x,t) \zeta(x',t') \rangle =2 \gamma  T
\delta(x-x')\delta(t-t')$.

\section{Static kink}
At first we construct the  static kink for $F=0$, which is a
localized metastable excitation. 
The static kink $z_k(x)$ 
is defined as the configuration that 
 minimizes the energy (\ref{hamil}), 
i.e., is a time-independent solution of (\ref{EOM})  in the absence of 
thermal noise ($\zeta =0$)
for   boundary conditions
$z_{k} (\pm L/2) = \pm a$ and 
  $\partial_x z_{k}|_{\pm L/2}=0$. 
For $F=0$ the potential is symmetric and 
$V(z)=V(-z)$ such that the kink configuration  is anti-symmetric
with 
$z_k(x)=-z_k(-x)$ and  centered at
$x=0$ (i.e.\ $z_k(0)=0$). 
For our  
 piecewise defined potential we  have to fulfill  five matching  conditions 
at $x=0$ which connect the two parts $x<0$ and $x>0$ of the kink:
 $z_{k}(-0) = z_{k}(+0)=0$,  
$\partial_x^{n} z_{k}|_{-0}=\partial_x^{n}  z_{k}|_{+0}$ 
for $n=1,2,3$.
 Both  parts $z_k(x)+a$  for  $x<0$ and $z_k(x)-a$ for 
$x>0$ of the 
  static kink  are linear
combinations of the four functions 
$e^{{\pm}x/w_k}e^{{\pm}ix/w_k}$ where 
the  eight linear expansion coefficients  are determined from 
the   boundary and matching conditions. The width $w_k$ 
of the kink and the energy $E_k$ 
of a single static  kink in the thermodynamic limit 
of large  $L$ are given by 
\begin{equation}
 w_k = \sqrt{2}x_{sc} = \sqrt{2}(\kappa/V_0)^{1/4}
~~~\mbox{and}~~~
 E_{k} = E_{sc}/\sqrt{2} = a^2 \kappa^{1/4} V_0^{3/4}/\sqrt{2}
~.
\label{wk}
\end{equation}
We expect our results for the  kink energy
$E_k{\sim}E_{sc}$ and width $w_k{\sim}x_{sc}$ to hold for all
 potentials with a barrier
height $\sim V_0a^2$ and potential minima separation $\sim a$
independent of the particular potential form; 
only numerical prefactors will differ. 
We want to point out that measurements of  the
kink width $w_k$ and the critical force density 
$F_c$ or the kink energy $E_k$  are sufficient to determine 
 the bending rigidity  $\kappa = F_cw_k^{4}/4a= E_kw_k^3/2a^2$ 
and thus the
persistence length $L_p =  2\kappa/T$ if the distance  $2a$ between
potential minima is known.

A static single kink  in a polymer of length $L$ is equivalent 
to one half of  a symmetric 
kink-antikink pair configuration
with  kink-antikink separation $d=L$
in a polymer of length  $2L$, as shown in fig.~\ref{potential}.
The kink-antikink interaction energy 
$E_{int}(d)=2(E_k(d)-E_k(\infty))$ can thus be found by determining 
the single  kink energy in a polymer  of length $L=d$.
For large separation $d/w_k \gg 1$ we find an exponential decay
$E_{int}(d) \sim \exp(-d/w_k)$.

A semiflexible polymer will stay localized to the 
potential wells even if we set $V(z)=0$ for $|z|>2a$ as long as $V_0>
V_{0,c}$ with 
$V_{0,c}a^2 \simeq (T/L_p)(L_p/a)^{2/3}$ according to \cite{KL03}. 
This condition is equivalent to $E_k \gg T$ and thus a small 
density of thermally induced kink excitations. 
A small kink density
 in combination with the condition $L_p \gg a$ 
also ensures that the semiflexible  polymer stays 
  oriented along the  $x$-axis such that the Hamiltonian 
(\ref{hamil}) stays  valid. 
The condition  $E_k \gg T$ of a  small kink density  is equivalent
to $L_p \gg w_k^3/a^2$. 
For sufficiently strong substrate potentials
 this  gives  a much wider range of
applicability of the Hamiltonian (\ref{hamil}) than in the 
absence of a potential  where 
 the condition $L_p > L$ of weak bending has to be fulfilled for a
semiflexible polymer to be oriented.

\section{Moving kink}
\begin{figure}[ht]
\begin{center}
  \epsfig{file=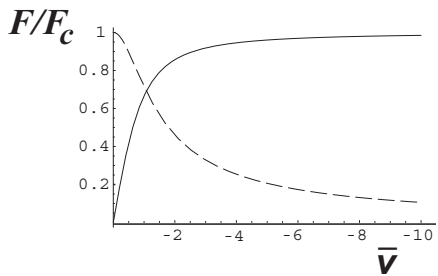,width=0.4\textwidth}
\caption{\label{Fveps}
Force density  $F$  (in units of $F_c$, solid line) and friction constant
$\eta_k$ 
(in units of $3F_ca/2^{3/2}v_{sc}=3a^2\gamma/w_k$, dashed line)
as function of velocity $\bar{v}= 3^{3/4}v/4v_{sc}$ 
for a moving kink. 
}
\end{center}
\end{figure}

A driving force density $F$ leads to an asymmetry in the potential 
and an effective force on kinks. 
Moving a kink by $-\Delta x$  
increases the  polymer length in the lower
 potential minimum by 
 $\Delta x$ and  leads to an  energy gain $-2aF\Delta x$ and thus 
a constant force ${\cal F}_k = -2aF$ on a kink. As argued above deviations
from kink interactions are exponentially small for separations $d \gg
w_k$. The force ${\cal F}_k$ leads to kink motion 
 such that 
 we also have to consider  moving kink solutions. 
For constant kink velocity $v$ the kink configuration  assumes a form 
$z_k(x,t) = z_k(x-vt)$ that solves (\ref{EOM}) for $\zeta=0$. 
Introducing the   coordinate
$y \equiv x-vt$  for the comoving frame, equation   (\ref{EOM}) reduces to 
\begin{equation}
\kappa \partial^{4}_y z_k
   -v \gamma \partial_y z_k    + V'(z_k) =0
\label{noisetau}
\end{equation}
which has to be solved with boundary conditions
analogously to  the static kink. However, in the
asymmetric potential the kink is no
longer symmetric but  centered at $y_0 \neq 0$
with  $z_k(y_0)=0$ where we also have to evaluate the 
 matching conditions.
Eq.~(\ref{noisetau}) can be made
dimensionless by measuring time in units of a characteristic time 
$t_{sc}={\gamma}/V_0$ and velocities  in units of $v_{sc} 
= x_{sc}/t_{sc} = \kappa^{1/4}V_0^{3/4}/\gamma$.
For a moving kink both parts  $z_k(y)-z_{min}^{-}$  
for  $y<0$ and $z_k(y)-z_{min}^{+}$ for 
$y>0$   are linear
combinations of  four functions 
$e^{K_{n}y}$ where $K_n$ ($n=1,...4$) are the four roots of the 
equation $\kappa K_n^4 -v\gamma K_n + V_0 =0$ that real part of which 
determine the width of the kink $w_k(v) \sim 1/|Re(K_{n})|$.
We find $K_{n}w_k = \pm H^{1/2}(\bar{v}) \pm (- H(\bar{v})
\pm 2^{3/2}3^{-3/4}\bar{v}H^{-1/2}(\bar{v}))^{1/2}$ (the first
and third sign have to be  identical)
where $\bar{v} = 3^{3/4}v/4v_{sc}$ is a dimensionless velocity and 
$H(\bar{v}) = 3^{-1/2}((\bar{v}^2+\sqrt{\bar{v}^4-1})^{1/3}+
   (\bar{v}^2+\sqrt{\bar{v}^4-1})^{-1/3})$ 
an increasing, real function with $H(\bar{v})\ge H(0)=1$ and
$H(\bar{v}) \sim \bar{v}^{2/3}$ for $\bar{v}\gg 1$.
The width of the moving kink  decreases with velocity as 
 $w_k(\bar{v}) = w_k H^{-1/2}(\bar{v})$ (for $\bar{v} <1$).
In the limit of large polymer length $L \gg w_k(v)$  a moving kink 
solution, fulfilling all boundary and matching conditions, must 
satisfy  the 
force-velocity relation 
\begin{equation}
  F(\bar{v}) =  - F_c\bar{v} \frac{3^{1/4}2^{-1/2}H^{3/2}(\bar{v})}
             {H^{3}(\bar{v})+ 3^{-3/2}\bar{v}^2}
~.
\label{Fv}
\end{equation}
 For small force densities, we find a linear relation
$F\approx -3^{1/4}2^{-1/2}F_c\bar{v}$, 
close to the critical force density $F_c$
the velocity diverges as $-\bar{v} \sim (1-F/F_c)^{-3/2}$, see
fig.~\ref{Fveps}.

The result (\ref{Fv}) can also be used to obtain the friction constant
${\eta}_k$ of a moving kink by equating the friction force $v\eta_k$
with the driving force ${\cal F}_k=-2aF$ which  
gives the relation $\eta_k= 2a|F(v)|/v$, see fig.~\ref{Fveps}. 
${\eta}_k$ is also related to the 
energy dissipation rate $dE/dt$ due to kink motion which is
defined as  the product
of the friction force $-v{\eta}_k$ and velocity: $dE/dt=-v^2{\eta}_k$.
On the other hand, $dE/dt$ can be calculated directly using the
equations of motion (\ref{EOM}) and (\ref{noisetau}) in the limit 
of large $L$
\begin{equation}
 \frac {dE}{dt} = \int_{-\infty}^{+\infty}dx
  \frac{\delta {\cal H}}{\delta z_k} 
  \partial_t z_k
  =-{\gamma}v^2 \int_{-\infty}^{+\infty}dy
 \left( \partial_y z_k \right)^2
~,
\label{energydissipation}
\end{equation}
and we read off a kink friction constant
 $\eta_k={\gamma}\int_{-\infty}^{+\infty}dy  \left(\partial_y
 z_k\right)^2$.
Integration in the limit of small driving forces 
gives $\eta_k \approx 3\gamma a^2/2w_k$ and 
equating the friction force
with the driving force $-2aF= v\eta_k$ 
gives a linear relation $v=- 4Fw_k/3\gamma a$ which agrees to leading
order with our above result (\ref{Fv}), see also fig.~\ref{Fveps}.

\section{Thermal noise and kink motion}
For a more detailed analysis of the effect of noise on the kink
motion we 
consider noise-induced perturbations of shape and velocity 
 of  a kink moving with constant velocity $v$.
For  a time-dependent kink center at $x_k(t)$ the comoving frame coordinate
is given by $\bar{y}\equiv x-x_k(t)$. 
Adding shape perturbations to the corresponding 
kink solution $z_k(\bar{y})$ of (\ref{noisetau}), we arrive at 
the decomposition
\begin{equation}
z(x,t)=  z_k(x-x_k(t))+
       \sum_{p=1}^{\infty} X_p(t) \phi_p(x-x_k(t),t)
.
\label{kinkposition}
\end{equation}  
$\phi_p$ are  normal modes
of the kinked polymer  which we will determine below
 and $X_p(t)$ are expansion coefficients; the zero mode of
kink translation is explicitly taken into account by positioning the 
 kink center at $x_k(t)$. 
Substituting   (\ref{kinkposition}) into the equation of motion 
(\ref{EOM}) and 
expanding about  the kink, 
 we obtain 
\begin{eqnarray}
 \gamma (v-{\dot x}_k)\left( \partial_{\by} z 
    + \sum_{p=1}^{\infty} X_p \partial_{\by} \phi_p \right)
 + \gamma \sum_{p=1}^{\infty}{\dot{X}_p} \phi_p
=\zeta
\label{field}
\end{eqnarray}
 if the normal modes 
 $\phi_p(\by,t)= f_p(\by)e^{-{\omega}_p t}$ 
fulfill the eigenvalue equation
\begin{equation}
 \kappa
\partial_{\by}^4 f_p  - \gamma v \partial_{\by} f_p 
  + V''(z_k(\by)) f_p = \omega_p \gamma f_p
\label{noiseeq}
\end{equation}
where $V''(z)=V_0(1-2a{\delta}(z))$. 
(\ref{noiseeq}) has to be solved with boundary conditions
$f_p(-L/2)=f_p(L/2)=0$ and  $f_p'(-L/2)=f_p'(L/2)=0$ where we consider  
the limit $L/2 \gg x_k(t)$ and  
 neglect the shift of boundaries in the comoving frame. 
Eq.~(\ref{noiseeq}) has a set of eigenvalues ${\omega}_p$ with 
orthonormal eigenfunctions 
$f_p(\by)$ (with respect to the scalar product 
$\langle f|g \rangle \equiv \int d\by f(\by) g(\by)$).
The translation mode  $f_0= \partial_{\by} z_k(\by)/C$ of the kink
has zero  eigenvalue $\omega_0=0$.
 $C$ is a normalization constant determined by 
 $C^2= \langle \partial_{\by} z_k | \partial_{\by} z_k \rangle$.
Multiplying eq.~(\ref{field}) with the 
translation mode $f_0$  and integrating 
yields an equation of motion for the kink
\begin{equation}
{\dot x}_k =v+
 \zeta_k
 \left[ 1+ C^{-1} \sum_{p=1}^{\infty} X_p
  e^{-\omega_p t} 
 \langle f_0|{\partial}_{\by}f_p \rangle \right]^{-1}
\label{adot}
\end{equation}
where  
$\zeta_k(t)= -(C\gamma)^{-1} \int d\by f_0(\by)
\zeta(\by+x_k(t),t)$ is an   effective Gaussian 
thermal  noise for the kink with correlations
 $\langle \zeta_k(t)\zeta_k(t')\rangle= \delta(t-t')(2
T/C^2\gamma)$
(where we used $\langle f_0|f_0 \rangle=1$). 
The sum in (\ref{adot}) represents terms from 
 kink-phonon scattering neglecting of which leads to an 
overdamped Langevin equation
${\dot x}_k(t)= v+\zeta_k(t)$ describing Brownian motion 
with drift.
 From the noise correlations 
 we can read off the corresponding 
 diffusion constant of the kink as
$D_k = T/C^2\gamma$.
Note that the corresponding kink friction constant 
$\eta_k = T/D_k$
 is identical to our above result (\ref{energydissipation})
 obtained from complementary energetic considerations.

If kink-phonon scattering is neglected, the kink is 
performing a Brownian motion with drift.
The polymer crosses the potential  barrier by moving a kink 
over the entire length $L$ of the polymer. Thus, the average crossing time 
is $t_{cr} \sim L/v$ for the case of directed diffusion  with $v>0$
under the influence of a driving force density $F$. 
For  $F=0$ and  $v=0$
the kink performs an unbiased random walk 
with $\langle x_k^2 \rangle \sim D_k t$ 
from which we estimate 
the average crossing time as $t_{cr} \sim L^2 \eta_k/ T
\sim L^2 \gamma a^2/ Tw_k$.
For $F=0$ and at low temperatures  
 $t_{cr}$ gives the relaxation time from a 
    kinked state as in fig.~\ref{potential}  to a kinkless state. 
The diffusive part of the  kink motion can be neglected for 
  forces  $F \gg 2T/ La$.

\section{Thermally activated barrier crossing}

\begin{figure}[t]
\begin{center}
  \epsfig{file=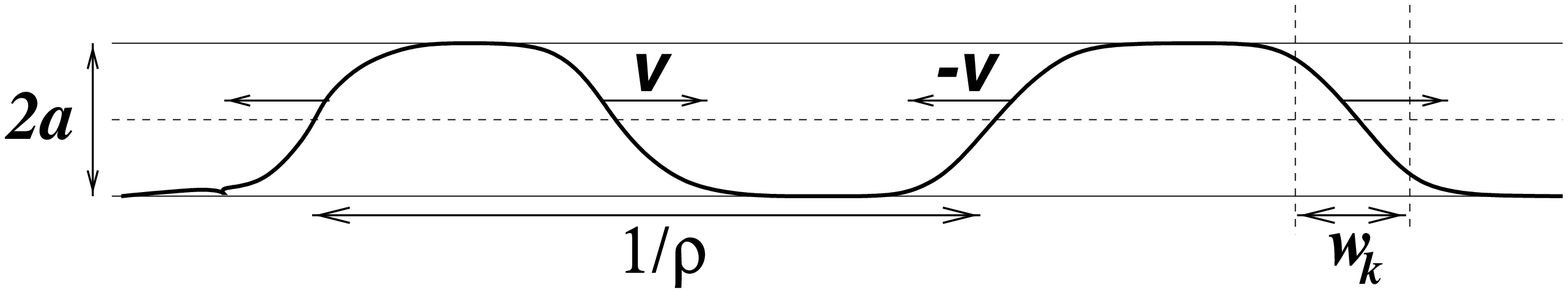,width=0.7\textwidth}
\caption{\label{meanvel}
An ensemble of well-separated kinks and antikinks which move with velocity
$v$ and $-v$, respectively.
}
\end{center}
\end{figure}

For sufficiently large $F$, 
thermally activated barrier crossing proceeds by 
the nucleation and subsequent separation of a kink-antikink pair, 
see fig.~\ref{meanvel}. 
Each passing kink or antikink increases the polymer position by
$\Delta z=2a$. 
For an ensemble of $\rho L$ kinks and  $\rho L$ antikinks with  kink density
${\rho} \ll 1/w_k$ as in fig.~\ref{meanvel}, 
 the fraction of moving polymer segments is given by 
$2 \rho L\, w_k/L=2\rho w_k$. These polymer segments
 move with velocity $2av/w_k$ in the $z$-direction which leads to the average 
velocity $v_z \equiv\langle \partial_t z\rangle =4av{\rho}$. 
The kink density $\rho$ is determined by the
dynamical equilibrium of kink nucleation with rate $j$ (per length) 
that we will calculate below, see (\ref{j}), 
and kink-antikink recombination with rate  $2\rho^2 v$ \cite{BL79}. 
Equating both rates gives a steady-state density $\rho^2=j/2v$ and 
thus an average polymer velocity
$v_z =2a(2vj)^{1/2}$.

In order to find the nucleation rate $j$,  we use Kramers theory. 
In the following,  we only give the main results of this 
calculation, details  will be described elsewhere.
As for   flexible strings \cite{BL79},  the dynamics of the nucleation
is  governed by the  critical nucleus representing 
the  saddle point in the multi-dimensional energy landscape. 
The critical nucleus configuration  $z_n(x)$ 
is the analogon of the static kink-antikink pair for $F>0$ and 
  fulfills the same saddle-point equation 
${\delta {\cal H}}/{\delta z}=0$, see (\ref{EOM}). 
For  the critical nucleus we obtain an excess energy
$\Delta E_n \approx 2E_k(1-F/F_c)^2$ that enters  
the   nucleation current $j \sim \exp(-\Delta E_n/T)$. 
The prefactor depends on  the  corresponding
attempt frequencies and, thus, the spectra  $\omega_{n,p}$ and
$\omega_{s,p}$ of fluctuations around the 
 the critical nucleus and the straight polymer, respectively. 
For the straight  polymer we  find a
spectrum of stable phononic  modes  $\omega_{s,0}=V_0/{\gamma}$ and
$\omega_{s,p}\approx V_0/\gamma
 + \kappa((\pi/2+p\pi)/L)^4/\gamma$ ($p\ge 1$). 
For the critical nucleus, the spectrum consists of an unstable mode
$\omega_{n,0}<0$ representing the collective coordinate along which 
the nucleation proceeds, a zero  mode $\omega_{n,1}=0$ corresponding to
the translation of the nucleus, and a sequence of
stable phononic modes $\omega_{n,2}=V_0/{\gamma}$ and 
$\omega_{n,p} \approx V_0/{\gamma}+ \kappa((b+p{\pi})/L)^4/\gamma$ ($p
\ge 3$),
where $b$ is a numerical constant. 
Using  Kramers theory in the regime  $F> T/2aw_k$ \cite{HMS},  
we finally obtain the nucleation rate
\begin{equation}
j= (2\pi)^{-3/2} \gamma^{1/2} G T^{-1/2} Q_n \exp\left(-\Delta E_n/T\right)
\label{j}
\end{equation}
where $Q_n^2 \equiv |\omega_{n,0}|\omega_{s,0}\omega_{s,1}
\prod_{p>1}\omega_{s,p}/\omega_{n,p}\approx  
|1-2^{4/3}(1-F/F_c)^{-8/3}|(V_0/\gamma)^3$
contains  all attempt frequencies, and 
 $G \equiv  L^{-1}\int dx_n [ \int dx\left(
\partial_x z_n(x)\right)^2]^{1/2}
\approx a(1-F/F_c)/\sqrt{w_k}$ is the Jacobian for the change of
coordinates from the amplitude of the translational mode 
$\partial_x z_n$ to the nucleus position $x_n$.

 For  small driving force densities  $F \ll 2{\rho}T/a$,
the  kink motion is  
diffusive,  and the above approach breaks down as kink-antikink pairs 
cannot separate but tend to recombine. 
For $E_k \gg T$, the system
 reaches thermodynamic equilibrium 
with a low kink density  
$\rho_{eq} \sim  \exp\left(-E_k/T\right)$
 given by the Boltzmann distribution and 
with $v_z=4a v \rho_{eq}$. 
For intermediate driving forces $2{\rho}T/a \ll F < F/2aw_k$,
the critical nucleus is  in quasi-equilibrium \cite{HMS}, and  we find 
again $j \sim \exp(-\Delta E_n/T)$ as in the high-force expression 
(\ref{j}) but with a different parameter dependence of the prefactor.

\section{Conclusion}

In conclusion, we  described the activated dynamics of semiflexible
polymers which is governed by kink excitations. 
The static kink has the energy $E_k$  
and the width $w_k$ as given by (\ref{wk}). 
Both kink properties are governed by the bending rigidity of the
semiflexible polymer.
In the presence of a driving force density $F$, 
there is a force ${\cal F}_k$ acting on the kink that leads to moving 
kink solutions 
which satisfy the force-velocity relation
 (\ref{Fv}). In the absence of kink-phonon
scattering the kink performs Brownian motion with drift for which we
have calculated 
the friction constant $\eta_k$ and the diffusion constant $D_k$. 
 This leads to estimates for the crossing times 
$t_{cr}\sim L/v$ for large forces $F\gg 2T/La$ and  
$t_{cr}\sim L^2\eta_k / T$ for small forces $F\ll 2T/ La$.
For large forces, the nucleation of kinks proceeds by activation 
over the saddle point which represents  the critical nucleus. 
Application of Kramers
theory leads to the nucleation rates (\ref{j})
which determine the average velocity 
$\langle \partial_t  z\rangle$ of the polymer. 
Our results are not only relevant to the dynamics of semiflexible
polymers but can be extended to kink excitations in fluid membranes 
\cite{AL96}.



\begin{thebibliography}{99}
  
\bibitem{Kramers} 
 \Name{Kramers H. A.}
   \REVIEW{Physica (Utrecht)}{7}{1940}{284}.
  
\bibitem{Haenggi} 
  \Name{H\"anggi P., Talkner P., \and Borkovec M.}
    \REVIEW{Rev. Mod. Phys.}{62}{1990}{251}.
  
\bibitem{dislocation} 
 \Name{Hirth J. P. \and Lothe J.}
 \Book{Theory of Dislocations} 
 \Publ{McGraw-Hill, New York}
 \Year{1968}.

\bibitem{vortex} 
 \Name{Blatter G., Feigelman M. V., Geshkenbein V. B., 
   Larkin A. I., \and Vinokur V. M.}
  \REVIEW{Rev. Mod. Phys.}{66}{1994}{1125}.


\bibitem{CDW} 
 \Name{Rice M. J., Bishop A. R., Krumhansl J. A., \and 
  Trullinger S. E.}
   \REVIEW{Phys. Rev. B}{36}{1976}{432}.
  
\bibitem{BL79} 
 \Name{B{\"u}ttiker M. \and Landauer R.}
 \REVIEW{Phys. Rev. Lett.} {43}{1979}{1453}; 
  \REVIEW{Phys. Rev. A}{23}{1981}{1397}.

\bibitem{CKBT80} 
  \Name{Currie J. F., Krumhansl J. A., Bishop A. R., \and 
   Trullinger S. E.}
  \REVIEW{Phys. Rev. B} {22}{1980}{477}.  

\bibitem{sebastian} 
 \Name{Sebastian K. L.} 
   \REVIEW{Phys. Rev. E}{61}{2000}{3245};
 \Name{Sebastian K. L. \and Paul A. K. R.}
  \REVIEW{Phys. Rev. E}{62}{2000}{927}.

\bibitem{taylor90} 
  \Name{Taylor W. H. \and Hagerman P. J.}
    \REVIEW{J. Mol. Biol.}{212}{1990}{363}.
  
\bibitem{kaes94} 
  \Name{K\"as J., Strey H., and Sackmann E.}
    \REVIEW{Nature}{368}{1994}{226}.


\bibitem{gittes93} 
 \Name{Gittes F., Mickey B., Nettleton J., \and Howard J.}
\REVIEW{J. Cell Biol.}{120}{1993}{923}.


\bibitem{NJ99}
 \Name{Netz R. N. \and Joanny J.-F.}
  \REVIEW{Macromolecules}{32}{1999}{9013}.

\bibitem{GB89}
 \Name{Gompper G. \and   Burkhardt T. W.}
   \REVIEW{Phys. Rev. A}{40}{1989}{6124}.


\bibitem{KL03} 
 \Name{Kierfeld J.  \and Lipowsky R.}
  \REVIEW{Europhys. Lett.}{62}{2003}{285}.


\bibitem{electrophoresis} 
   \Name{Han J.,  Turner S. W., \and  Craighead H. G.}
  \REVIEW{Phys. Rev. Lett.}{83}{1999}{1688}.


\bibitem{kurth02} 
  \Name{Kurth D. G.,  Severin N., \and  Rabe J. P.}
    \REVIEW{Angew. Chem.}{114}{2002}{3833}.

\bibitem{CM01}
 \Name{Costantini G. \and Marchesoni F.}
 \REVIEW{Phys. Rev. Lett.}{87}{2001}{114102}.

\bibitem{HMS}
 \Name{H{\"a}nggi P., Marchesoni F., \and  Sodano P.}
\REVIEW{Phys. Rev. Lett.}{60}{1988}{2563}.


\bibitem{AL96}  
 \Name{Ammann A. \and Lipowsky R.}
   \REVIEW{J. Phys. II France}{6}{1996}{255}.

\end{thebibliography}
\end{document}